\documentclass[12pt]{article}
\linethickness{0.4pt}
\titlepage
\newcommand{\be}{\begin{equation}}
\newcommand{\ee}{\end{equation}}
\newcommand{\bea}{\begin{eqnarray}}
\newcommand{\eea}{\end{eqnarray}}
\author{Oscar Diego\footnote{EMail: odiego@apiaxxi.es}
\footnote{Present Address: Apia XXI, Departamento de Topograf\'\i a, 
Luis Mart\'\i nez 21, 39005 Santander, Spain}}
\title{Integral representation of the Ising model}

\begin{document}
\maketitle
\abstract{ The partition function of the 2D Ising model coupled to an external 
magnetic field is studied. We show that the sum over the spin variables can be 
reduced to an integration over a finite number of variables. This integration 
must be performed numerically. But in order to reduce the 
partition function we must introduce as many different coupling constants as 
spin variables. The total memory that we need in order to store these coupling 
constants imposed important restrictions on the number of spin variables. }
\newpage

{\bf 1. Introduction}

Physical models in Statistical Mechanics and Quantum Field Theory are
defined by integrals or sums over an infinite number of variables like:
\begin{equation}
\lim_{N \rightarrow \infty} \sum_{\sigma_{1}} \cdots \sum_{\sigma_{N}}
\exp{ \{W(\sigma_{1}, \cdots , \sigma_{N})\}} .
\label{eq:1.1}
\end{equation}
The weight $W$ is the energy in Statistical Mechanics and the classical
action in Quantum Field Theory. The variables $\{ \sigma_{i} \}$ are
coupled through terms like
\begin{equation}
\sum_{i} \sigma_{i} \sigma_{i+1} .
\label{eq:1.2}
\end{equation}

In general we cannot solve exactly ($\ref{eq:1.1}$). Actually these sums
can be solved exactly only if they can be transformed
into sums over uncoupled variables. For instance, in free field models
the Fourier coefficients of the field variables are not
coupled.
In Statistical Mechanics the 1D Ising model{\cite{1}} and the 2D Ising model
without magnetic field are equivalent to free fermion
models \cite{2,3}.

In this paper we are going to show that we can decouple the
spin variables of the partition
function of the 2D Ising
model with an homogeneous magnetic field. This is an interesting
result because the 2D Ising model with magnetic field cannot
be solved exactly.

In order to decouple the spin variables we are going to use the following 
trick. Let us remark that if the interacting term between the variables is given by
\begin{equation}
\left ( \sum_{i} \sigma_{i} \right )^{2} ,
\label{eq:1.3}
\end{equation}
then we can decouple the variables using the following
identity
\begin{equation}
\int d x \exp {\left \{ -x^{2} + 2 x \sum_{i} \sigma_{i} \right \}} \propto
\exp { \left \{ \left ( \sum_{i} \sigma_{i} \right )^{2} \right \}} .
\label{eq:1.4}
\end{equation}

Let us consider
\begin{equation}
\lim_{T \rightarrow \infty} \frac{1}{T} \int_{0}^{T} d \omega
\left ( \sum_{j} e^{i \omega \theta_{j}} \sigma_{j} \right ) 
\left ( \sum_{k} e^{-i \omega \theta_{k-1}} \sigma_{k} \right ) , 
\label{eq:1.5}
\end{equation}
where $\{ \theta_{j} \}$ is a set of different constants.
In other words
\bea
\theta_{i} \neq \theta_{j} & if & i \neq j .
\eea

Let us remark that
\be
\lim_{T \rightarrow \infty} \frac{1}{T} \int_{0}^{T} d \omega
e^{i \omega \theta} \neq 0
\ee
only if $\theta$ is zero. Hence ($\ref{eq:1.5}$) becomes
\be
\sum_{j} \sum_{k} \delta (\theta_{j}, \theta_{k-1}) \sigma_{j} \sigma_{k} .
\ee
But the constants $\{ \theta_{i} \}$ are different. Hence the above
expression becomes
\be
\sum_{j} \sum_{k} \delta_{j,k-1} \sigma_{j} \sigma_{k-1} =
\sum_{j} \sigma_{j} \sigma_{j+1} .
\ee

Hence we can represent the interacting term
\be
\sum_{i} \sigma_{i} \sigma_{i+1}
\ee
as a double sum over independent indices and we can use $(\ref{eq:1.4})$ in 
order to decouple the spin variables.

In the next section we will apply this trick to the 2D Ising model with a
magnetic field.

Once the spin variables are decoupled we can integrate over all the
spin variables. Hence we will show that the partition function
of the 2D Ising model is given by an integral over a finite
number of variables.

From a theoretical point of
view this result is very interesting because the 2D Ising model with
magnetic field cannot be solved exactly but we will show that
the sum over the arbitrary number of spin variables can be
reduced to an integral over a finite number of variables.

But from a practical point of view the trick developed in this paper
has an important drawback. We will show that the set of coupling
constants $\{ \theta_{i} \}$ must take their values on a very wide
range of numbers. Each coupling constant $\theta_{i}$ has a fixed value
but if the number of spin variables is large then some of the
coupling constants must take very large values. Therefore the
number of spin variables cannot be large.

Therefore this approach can only be used for small lattices. But with the 
hypothesis of finite-size scaling\cite{4} we do not need very large lattice in 
order to study the physics of the model at the thermodynamic limit. 
Moreover for some boundary conditions 
the model reaches the thermodynamic limit faster\cite{5,6}. 

In\cite{6} it has been shown that the dimer approach\cite{7,8} can be used in 
numerical calculation for small lattices. But the dimer approach cannot be 
used when a magnetic field is present or in 3D. Actually the approach that we study 
in this paper can be generalized to the 3D Ising model.

{\bf 2. The partition function of the 2D Ising model}

Let us consider the set $\{ \sigma_{i} \}$ of spin variables:
\be
\sigma_{i} = \pm 1
\label{eq:2.1}
\ee
defined over vertices of the square lattice. Index $i$ labels columns
and $j$ labels rows. Let us define the energy:
\be
E = \sum_{i,j} \beta \sigma_{i,j} ( \sigma_{i,j+1} + \sigma_{i+1,j} )
+ \alpha \sigma_{i,j} .
\ee
The 2D Ising model with coupled to an external magnetic field is defined
by the partition function:
\be
Z = \sum_{\sigma_{i,j}} \exp{ [ E ] } .
\label{eq:Part.Funct.1}
\ee
Because the spin variables are defined by ($\ref{eq:2.1}$) the partition
function is also given by
\bea
Z & = & (\cosh{\beta})^{N_{L}} (\cosh{\alpha})^{N_{V}}
\sum_{\sigma_{i,j}} \prod_{i,j} \nonumber \\
 & & ( 1 + z \sigma_{i,j} \sigma_{i,j+1})
( 1 + z \sigma_{i,j} \sigma_{i+1,j})
( 1 + h \sigma_{i,j} ) ,
\eea
where $N_{L}$ is the number of links and $N_{V}$ is
the number vertices of the lattice. And
\bea
z & = & \tanh{\beta} \nonumber \\
h & = & \tanh{\alpha} .
\eea
For each $\sigma_{i,j}$ let us define four new spin
variables: $\sigma^{U}_{i,j}$, $\sigma^{D}_{i,j}$, $\sigma^{R}_{i,j}$
and $\sigma^{L}_{i,j}$. Now let us define the partition function:
\bea
\tilde{Z} & = A \sum_{\sigma} \prod_{i,j} &
( 1 + \sqrt{z} \sigma_{i,j} \sigma^{U}_{i,j} )
( 1 + \sqrt{z} \sigma_{i,j} \sigma^{D}_{i,j} ) \nonumber \\
 & & ( 1 + \sqrt{z} \sigma_{i,j} \sigma^{L}_{i,j} )
( 1 + \sqrt{z} \sigma_{i,j} \sigma^{R}_{i,j} ) \nonumber \\
 & & ( 1 + h \sigma_{i,j} )
( 1 + \sigma_{i,j}^{R} \sigma^{L}_{i+1,j} )
( 1 + \sigma_{i,j}^{D} \sigma^{U}_{i,j+1} ) ,
\label{eq:ztil}
\eea
where
\be
A = \left ( \frac{1}{2} \right )^{4 N_{V}}
\left ( \cosh{\beta} \right )^{N_{L}} 
\left ( \cosh{\alpha} \right )^{N_{V}} .
\ee
It is very easy to show that
\bea
& \sum_{\sigma_{i,j}^{R} = \pm 1} 
\sum_{\sigma_{i+1,j}^{L} = \pm 1} & 
( 1 + \sqrt{z} \sigma_{i,j} \sigma^{R}_{i,j} )
( 1 + \sigma_{i,j}^{R} \sigma^{L}_{i+1,j} ) \nonumber \\
& & ( 1 + \sqrt{z} \sigma^{L}_{i+1,j} \sigma_{i+1,j}  )  \nonumber \\
& & = 2^{2} ( 1 + z \sigma_{i,j} \sigma_{i+1,j} ) .
\eea
Hence $Z$ and $\tilde{Z}$ are equal.

Now let us define the following set of constants $\{ \theta_{i,j} \}$.
\bea
\theta_{1,1} & = & 1 \nonumber \\
&\cdots&  \nonumber \\
\theta_{1,n+1} & = & \sum_{k=1}^{n} \theta_{1,k}+ 1 \nonumber \\
&\cdots& \nonumber \\
\theta_{m,n+1} & = & \sum_{j=1}^{m-1} \sum_{k=1}^{N} \theta_{j,k} +
\sum_{k=1}^{n} \theta_{m,k} + 1 .
\eea
The solutions of these equations are
\be
\theta_{m,n} = 2^{(m-1)N+n-1} .
\ee

Now let us consider the following linear combination
\be
\sum_{i=1}^{N} \sum_{j=1}^{N} n_{i,j} \theta_{i,j} = 0
\quad n_{i,j} = 0, \pm 1 .
\label{eq:2.comb}
\ee

We are going to show that $(\ref{eq:2.comb})$ holds only if the
integer coefficients $n_{i,j}$ are zero.
It is very easy to show that
\be
\theta_{N,N} > \arrowvert
\sum_{i=1}^{N} \sum_{j=1}^{N-1} n_{i,j} \theta_{i,j}
\arrowvert
\quad \forall n_{i,j} = 0 , \pm 1 .
\ee
Hence if $(\ref{eq:2.comb})$ holds then $n_{N,N}$ must be zero.

In the same way we can show that
\be
\theta_{N,N-1} > \arrowvert
\sum_{i=1}^{N} \sum_{j=1}^{N-2} n_{i,j} \theta_{i,j}
\arrowvert
\quad \forall n_{i,j} = 0 , \pm 1 .
\ee
Then $n_{N,N-1}$ must be zero.

In general we can show that
\be
\theta_{m,n} > \arrowvert
\sum_{i=1}^{m-1} \sum_{j=1}^{N} n_{i,j} \theta_{i,j} +
\sum_{j=1}^{n-1} n_{m,j} \theta_{i,j}
\arrowvert
\quad \forall n_{i,j} = 0 , \pm 1 .
\ee
Then $n_{n,m}$ must be zero.

Hence if the linear combination defined in ($\ref{eq:2.comb}$) is
zero then all the coefficients $n_{i,j}$ must be zero.

Now let us define the partition function:
\bea
Z^{\prime} & = & A \sum_{\sigma} \lim_{T_{H} \rightarrow \infty}
\lim_{T_{V} \rightarrow \infty} \frac{1}{T_{H} T_{V}}
\int_{0}^{T_{H}} d \omega_{H} \int_{0}^{T_{V}} d \omega_{V} \prod_{i,j}
    \nonumber \\
    & &
    \left ( 1 + \sqrt{z} \sigma_{i,j} \sigma_{i,j}^{R} \right )
    \left ( 1 + \sqrt{z} \sigma_{i,j} \sigma_{i,j}^{L} \right )
    \nonumber \\
    & &
    \left ( 1 + \sqrt{z} \sigma_{i,j} \sigma_{i,j}^{U} \right )
    \left ( 1 + \sqrt{z} \sigma_{i,j} \sigma_{i,j}^{D} \right )
    \nonumber \\
    & &
    \left ( 1 + h \sigma_{i,j}  \right )
    \nonumber \\
 & &
    \left ( 1 + e^{i \omega_{H} \theta_{i,j}} \sigma_{i,j}^{R} \right )
    \left ( 1 + e^{-i \omega_{H} \theta_{i-1,j}} \sigma_{i,j}^{L} \right )
    \nonumber \\
    & &
    \left ( 1 + e^{i \omega_{V} \theta_{i,j}} \sigma_{i,j}^{U} \right )
    \left ( 1 + e^{-i \omega_{V} \theta_{i,j-1}} \sigma_{i,j}^{D} \right ) ,
\label{eq:2.Zunc}
\eea
Let us consider periodic boundary conditions. In this case the boundary
conditions of the constants $\{ \theta_{i,j} \}$ are given by
\bea
\theta_{0,j} & = & \theta_{N,j} \nonumber \\
\theta_{i,0} & = & \theta_{i,N} .
\eea

We are going to show that this partition function is equal to $\tilde{Z}$.
Let us remark that
\be
\prod_{i} ( 1 + a_{i} ) = 1 + \sum_{i} a_{i} + \sum_{i<j} a_{i} a_{j}
+ \sum_{i<j<k} a_{i} a_{j} a_{k} + \cdots .
\ee
We are going to use this formula in order to expand the products:
\bea
& \prod_{i,j} & \left ( 1 + e^{i \omega_{H} \theta_{i,j}} \sigma_{i,j}^{R} \right )
    \left ( 1 + e^{-i \omega_{H} \theta_{i-1,j}} \sigma_{i,j}^{L} \right )
    \nonumber \\
    & & 
    \left ( 1 + e^{i \omega_{V} \theta_{i,j}} \sigma_{i,j}^{U} \right )
    \left ( 1 + e^{-i \omega_{V} \theta_{i,j-1}} \sigma_{i,j}^{D} \right ) .
\label{eq:2.prod}
\eea
Hence ($\ref{eq:2.prod}$) is given by a sum of terms like 
\be
\cdots + e^{i \omega_{H} (\sum_{i,j}^{\prime} n_{i,j} \theta_{i,j})}
\prod_{i,j}^{\prime} \prod_{J}^{\prime} \sigma_{i,j}^{J}
e^{i \omega_{V} (\sum_{i,j}^{\prime \prime} m_{i,j} \theta_{i,j})}
\prod_{i,j}^{\prime \prime} \prod_{K}^{\prime \prime} \sigma_{i,j}^{K}
+ \cdots ,
\label{eq:2.exp}
\ee
where indices $J$ and $K$ can take the values
\bea
J & = & R, L \nonumber \\
K & = & U, D.
\eea
Let us remark that $n_{i,j}$ and $m_{i,j}$ can take the values
\bea
n_{i,j} & = & 0 , \pm 1 \nonumber \\
m_{i,j} & = & 0 , \pm 1 .
\eea
The primes over the sum and product symbols in ($\ref{eq:2.exp}$) means that the 
indices do not take all their values.

For instance, the partition function $Z^{\prime}$ depends on $\theta_{i,j}$
through the factor
\bea
\left ( 1 + e^{i \omega_{H} \theta_{i,j}} \sigma_{i,j}^{R} \right )
\left ( 1 + e^{-i \omega_{H} \theta_{i,j}} \sigma_{i+1,j}^{L} \right )
\nonumber \\
=
\left ( 1 + e^{i \omega_{H} \theta_{i,j}} \sigma_{i,j}^{R}
+ e^{-i \omega_{H} \theta_{i,j}} \sigma_{i+1,j}^{L}
+ \sigma_{i,j}^{R} \sigma_{i+1,j}^{L} \right) .
\label{eq:2.factn}
\eea
Let us remark that first and fourth term in the right hand side of the above equation 
correspond to $n_{i,j}=0$. The second term is defined by $n_{i,j}=1$ and the 
third term is related with $n_{i,j}=-1$

Now we are going to perform the integration
over $\omega_H$ and $\omega_V$ and to take the limits in $T_H$ and $T_V$.
The only terms of the expansion ($\ref{eq:2.exp}$) that survive
to the limits in $T_H$ and $T_V$ are those with all the
coefficients $n_{i,j}$ and $m_{i,j}$ equal to zero.
Therefore factors like ($\ref{eq:2.factn}$) becomes
\be
\left ( 1 + \sigma_{i,j}^{R} \sigma_{i+1,j}^{L} \right) .
\ee
Hence $Z^{\prime}$ is also given by:
\bea
Z^{\prime} & = & A \sum_{\sigma} \prod_{i,j} 
    \left ( 1 + \sqrt{z} \sigma_{i,j} \sigma_{i,j}^{R} \right )
    \left ( 1 + \sqrt{z} \sigma_{i,j} \sigma_{i,j}^{L} \right )
    \nonumber \\
    & & 
    \left ( 1 + \sqrt{z} \sigma_{i,j} \sigma_{i,j}^{U} \right )
    \left ( 1 + \sqrt{z} \sigma_{i,j} \sigma_{i,j}^{D} \right )
    \left ( 1 + h \sigma_{i,j}  \right )
    \nonumber \\
    &  &
    \left ( 1 + \sigma_{i,j}^{R} \sigma_{i+1,j}^{L} \right )
    \left ( 1 + \sigma_{i,j}^{U} \sigma_{i,j+1}^{D} \right ) .
\eea
This is the partition function $\tilde{Z}$ given
in ($\ref{eq:ztil}$).

{\bf 3. Integral representation of the partition function}

We have shown that the partition function of the 2D Ising model
with a magnetic field is equivalent to the partition
function $Z^{\prime}$ given in ($\ref{eq:2.Zunc}$).
Hence $Z$ can be written as:
\be
Z = A \lim_{T_{H} \rightarrow \infty}
\lim_{T_{V} \rightarrow \infty} \frac{1}{T_{H} T_{V}}
\int_{0}^{T_{H}} d \omega_{H} \int_{0}^{T_{V}} d \omega_{V}
\sum_{\sigma} \prod_{i,j} K_{i,j} ,
\ee
where $K_{i,j}$ is given by:
\bea
K_{i,j} & = &
    \left ( 1 + \sqrt{z} \sigma_{i,j} \sigma_{i,j}^{R} \right )
    \left ( 1 + \sqrt{z} \sigma_{i,j} \sigma_{i,j}^{L} \right )
    \nonumber \\
    & & 
    \left ( 1 + \sqrt{z} \sigma_{i,j} \sigma_{i,j}^{U} \right )
    \left ( 1 + \sqrt{z} \sigma_{i,j} \sigma_{i,j}^{D} \right )
    \left ( 1 + h \sigma_{i,j}  \right )
    \nonumber \\
    &  & 
    \left ( 1 + e^{i \omega_{H} \theta_{i,j}} \sigma_{i,j}^{R} \right )
    \left ( 1 + e^{-i \omega_{H} \theta_{i-1,j}} \sigma_{i,j}^{L} \right )
    \nonumber \\
    &  & 
    \left ( 1 + e^{i \omega_{V} \theta_{i,j}} \sigma_{i,j}^{U} \right )
    \left ( 1 + e^{-i \omega_{V} \theta_{i,j-1}} \sigma_{i,j}^{D} \right ) .
\eea
Let us remark that $K_{i,j}$ depends only on the spin variables defined at 
the vertex $(i,j)$. Hence
\be
\sum_{\sigma} \prod_{i,j} K_{i,j} = 
\prod_{i,j} \sum_{\sigma_{i,j}} K_{i,j}.
\ee 
We can perform the integration over the spin variables and the partition function $Z$ 
is given by
\be
Z = A \lim_{T_{H} \rightarrow \infty}
\lim_{T_{V} \rightarrow \infty} \frac{1}{T_{H} T_{V}}
\int_{0}^{T_{H}} d \omega_{H} \int_{0}^{T_{V}} d \omega_{V}
\prod_{i,j} \bar{K}_{i,j} ,
\ee
where
\bea
\bar{K}_{i,j} & = & \sum_{\sigma_{i,j} = \pm 1}
\sum_{\sigma_{i,j}^{R} = \pm 1}
\sum_{\sigma_{i,j}^{L} = \pm 1}
\sum_{\sigma_{i,j}^{U} = \pm 1}
\sum_{\sigma_{i,j}^{D} = \pm 1} \nonumber \\
    &  & 
    \left ( 1 + \sqrt{z} \sigma_{i,j} \sigma_{i,j}^{R} \right )
    \left ( 1 + \sqrt{z} \sigma_{i,j} \sigma_{i,j}^{L} \right )
    \nonumber \\
    & & 
    \left ( 1 + \sqrt{z} \sigma_{i,j} \sigma_{i,j}^{U} \right )
    \left ( 1 + \sqrt{z} \sigma_{i,j} \sigma_{i,j}^{D} \right )
    \left ( 1 + h \sigma_{i,j}  \right )
    \nonumber \\
    &  & 
    \left ( 1 + e^{i \omega_{H} \theta_{i,j}} \sigma_{i,j}^{R} \right )
    \left ( 1 + e^{-i \omega_{H} \theta_{i-1,j}} \sigma_{i,j}^{L} \right )
    \nonumber \\
    & & 
    \left ( 1 + e^{i \omega_{V} \theta_{i,j}} \sigma_{i,j}^{U} \right )
    \left ( 1 + e^{-i \omega_{V} \theta_{i,j-1}} \sigma_{i,j}^{D} \right ) .
\eea

It is very easy to show that $\bar{K}_{i,j}$ is given by
\be
\bar{K}_{i,j} = 2^{5} \hat{K}_{i,j} ,
\ee
where
\bea
\hat{K}_{i,j} & = & 1 +
z^{2} e^{i( \omega_{V} \theta_{i,j} - \omega_{H} \theta_{i-1,j} )} +
z^{2} e^{i \omega_{V} ( \theta_{i,j} - \theta_{i,j-1} )}
\nonumber \\
        & + &
z^{2} e^{i \theta_{i,j} ( \omega_{H} + \omega_{V} ) } +
z^{2} e^{-i( \omega_{H} \theta_{i-1,j} + \omega_{V} \theta_{i,j-1} )}
\nonumber \\
        & + &
z^{2} e^{i \omega_{H} ( \theta_{i,j} - \theta_{i-1,j} ) } +
z^{2} e^{i( \omega_{H} \theta_{i,j} - \omega_{V} \theta_{i,j-1} )}
\nonumber \\
        & + &
z^{4} e^{i (\omega_{H} \theta_{i,j} + \omega_{V} \theta_{i,j}
            - \omega_{H} \theta_{i-1,j} - \omega_{V} \theta_{i,j-1})}
\nonumber \\
        & + &
z \omega \left ( e^{i\omega_{V} \theta_{i,j}} + e^{i\omega_{H} \theta_{i,j}} +
    e^{-i\omega_{V} \theta_{i,j-1}} + e^{-i\omega_{H} \theta_{i-1,j}}
    \right )
\nonumber \\
        & + &
z^{3} \omega e^{i (\omega_{H} \theta_{i,j} 
            - \omega_{H} \theta_{i-1,j} - \omega_{V} \theta_{i,j-1})}
\nonumber \\
        & + &
z^{3} \omega e^{i (\omega_{V} \theta_{i,j}
            - \omega_{H} \theta_{i-1,j} - \omega_{V} \theta_{i,j-1})}
\nonumber \\
        & + &
z^{3} \omega e^{i (\omega_{H} \theta_{i,j} + \omega_{V} \theta_{i,j}
            - \omega_{H} \theta_{i-1,j} )}
\nonumber \\
        & + &
z^{3} \omega e^{i (\omega_{H} \theta_{i,j} + \omega_{V} \theta_{i,j}
           - \omega_{V} \theta_{i,j-1})} .
\label{eq:3.1}
\eea
Hence the partition function $Z$ has the following representation:
\be
Z = 2^{5N^{2}} A \lim_{T_{H} \rightarrow \infty}
\lim_{T_{V} \rightarrow \infty} \frac{1}{T_{H} T_{V}}
\int_{0}^{T_{H}} d \omega_{H} \int_{0}^{T_{V}} d \omega_{V}
\exp{ [ \sum_{i,j} \tilde{K}_{i,j} ] } ,
\ee
where
\be
\tilde{K}_{i,j} = \ln {\hat{K}_{i,j}}
\ee
and $\hat{K}_{i,j}$ is given by ($\ref{eq:3.1}$)

Hence we have transform a sum over an arbitrary number of variables
into and integral over two variables, a double sum over two indices and
two limits.

{\bf 4. Conclusions}

We have shown that the partition function of the 2D Ising model coupled
to an external magnetic field can be represented by an integral over a
finite number of degrees of freedom. This result is very interesting
because the 2D Ising model with magnetic field cannot be solved exactly.
 
The remaining finite integration must be
performed numerically. 
But it is difficult to perform numerical calculation with this
representation of the partition function. 
The main problem is the total memory that we need in order to store
the coupling constants $\theta_{i,j}$.
If $\theta_{i,j}$ are stored as 32 bits integers then
\begin{equation}
- 2 ^{31} < \theta_{i,j} < 2^{31} .
\end{equation}
If they are stored as 64 bits floating point then the maximum value
of $\theta_{i,j}$ is
\begin{equation}
\theta_{i,j} < 2^{1000} .
\end{equation}
In $2D$ this bound means that
\begin{equation}
N < 30
\end{equation}
but in $3D$
\begin{equation}
N < 10
\end{equation}

{\bf Acknowledgments}

I thank Gloria Media for her illuminating influence in this paper.

\end{document}